\documentclass[dvipdfmx,aps,twocolumn]{revtex4}
\usepackage{amsmath,amssymb,graphicx}

\newcommand{\simj}{\stackrel{>}{_\sim}}
\newcommand{\simk}{\stackrel{<}{_\sim}}

\begin{document}
\title{Plasmon effect  on the  Coulomb  pseudopotential $\mu^*$ in the McMillan equation }

\author{Kazuhiro Sano, Mithuki Seo, and Kohji Nakamura}
\affiliation{Department of Physics Engineering, Mie University, Tsu, Mie 514-8507, Japan} 

\begin{abstract}
We  examine the  Coulomb pseudopotential $\mu^*$  in the McMillan equation applying to the  superconductivity of heavily doped  semiconductors.
Systematic calculation  using the first-principles calculation suggests   that  $\mu^*$  should be  considered as a  variable quantity depending on  carrier density $n$ in  semiconductors, although it is usually considered as  a constant  about 0.1.
To clarify  $n-$dependence of  $\mu^*$, we  solve the McMillan equation inversely for  $\mu^*$ by combining the result of the first-principles calculation  and that of  experiments.
It indicates  that  $\mu^*$ decreases with $n$ and  becomes   negative under  $n \sim 5 \times 10^{-21}[{\rm cm^{-3}}]$. 
This reduction  is explained by the effect of  plasmon  which  may play an important role in the  superconductivity of  low carrier systems such as heavily  doped  semiconductors.
\end{abstract} 

\maketitle
Since the superconductivity in heavily Born-doped diamond  was observed\cite{Ekimov}, much effort has been done to find the superconductivity of not only  diamond but   semiconductors such as Silicon and Germanium.\cite{Takano,Klein,Kawano,Okazaki,Marcenat,Grockowiak,Herrmannsdorfer,Herrmannsdorfer2,Koonce,Kriener} 
If the BCS theory\cite{BCS} can be applied to these systems,  the transition temperature of superconductivity, $T_{\rm c}$, may be roughly described as $T_{\rm c} \sim \omega_{\rm ph} \exp(-\frac{1}{V N(E_{\rm F})})$,
where $\omega_{\rm ph}$ is a characteristic phonon  frequency of the system, $V$ is an effective attraction  between electrons, and  $N(E_{\rm F})$ is the density of state at the Fermi energy.
Because higher  $\omega_{\rm ph}$ may  be  expected to lead higher $T_{\rm c}$, 
 the superconductivity of  diamond has attracted much interest due to  its high  phonon  frequency  up to about 2000K  as the Debye  frequency  $\omega_{\rm D}$. 
Indeed, experiments\cite{Takano,Klein,Kawano,Okazaki} show that $T_{\rm c}$ reaches  about 10K or more at carrier  density,  $n \sim  \times 10^{22} {\rm cm^{-3}}$, which is corresponding to boron concentration $\rho \sim 8\%$.
Recent progress in implant technology of dopant also enable  to produce the  heavily doped  Si and Ge    with carrier  concentration up to several percent and to observe its superconductivity of $T_{\rm c} \sim$ 0.7K in boron-doped  Si\cite{Marcenat,Grockowiak}  and 1K in gallium-doped  Ge.\cite{Herrmannsdorfer,Herrmannsdorfer2}.  

In theoretical point of view, many works using the first-principles calculation address  the superconductivity of  semiconductors by assuming  a  traditional phonon mechanism.\cite{Boeri,Xiang,Ma,Moussa,Subedi,Sano}
In these theories,  so-called the McMillan equation\cite{McMillan,Allen} is used to estimate   $T_{\rm c}$ as a more sophisticated formula than the BCS theory,
\begin{align} 
T_{\rm  c} \simeq \frac{\omega_{\rm  log}}{1.2}\exp(-\frac{1.04(1+\lambda)}{\lambda-\mu^*(1+0.62\lambda) }),
\label{M_Tc}
\end{align} 
where $\omega_{\rm log}$ is a logarithmic average  frequency, $\lambda$ is the electron-phonon coupling constant, and $\mu^*$ is the  Coulomb pseudopotential.
Although the  equation  is produced as a semi-empirical formulation to fit the numerical solution of the Elashberug equation, estimated $T_c$ is well consistent with  many experiments of usual metals.

Since  it is not easy to determine  $\mu^*$ within the first-principles calculation,
its value  is usually  treated as a constant about 0.1, based on the work of  Morel and Anderson.\cite{Morel} 
They drive  $\mu^*$ within  Thoms-Fermi approximation    
\begin{align} 
\mu^*  = \frac{\mu}{1+\mu \ln(E_{\rm F}/\omega_{\rm D})},
\label{mu_s_Morel}
\end{align} 
where  $E_{\rm F}$ is the Fermi energy and  $\mu$ is the screened Coulomb potential 
\begin{align} 
\mu  = \frac{q_{\rm TF}^2}{ 8p_{\rm F}^2 }  \ln(1+\frac{4p_{\rm F}^2 }{ q_{\rm TF}^2} ).
\label{mu_Morel}
\end{align} 
Here,  $q_{\rm TF}$ is the Thomas-Fermi wave number and $p_{\rm F}$ is the Fermi momentum.
It is interpreted that $\mu$  is  renormalized to $\mu^*$  by the difference of the energy scale between phonon and   Coulomb potential.   
In usual metals, a typical value of   $\mu$ is 0.2 and   $\ln(E_{\rm F}/\omega_{\rm D}) \sim 5$, and then $\mu^*$ is roughly given by $\sim 0.1$.
When we adapt the above  $\mu^*$ into  Eq.\ref{M_Tc}, we can  estimate  a reasonable value of  $T_{\rm c}$ for many realistic materials. Because the first-principles calculation gives us   accurate values of  $\lambda$ and $\omega_{\rm log}$ without ambiguity.
It even succeeds in prediction of  new phonon mediated high-temperature superconductors such as SH$_3$\cite{Li} and LaH$_{10}$\cite{Liu} at high pressure. Therefor,  the above formulation might seem to  give us a useful recipe to estimate   $T_{\rm c}$ of  not only usual metal but also  heavily doped semiconductors. 

However, comparison   of theory and  experiment for semiconductors seems to be not well, although many theoretical works  analyze  $T_{\rm c}$  using  the  McMillan equation.
Especially, $n$-dependence of $T_{\rm c}$  can not been  explained at all.\cite{Bustarret} 
If we wish to explain  $n$-dependence of $T_{\rm c}$ in the framework of the  McMillan formulation, we should assume  $\mu^*$ not  a constant but a variable depending on  $n$.
In the case of diamond superconductor, Klein et al. have given   $\mu^*$ as a function of $n$ concretely.
They  substitute  experimental result of  $T_{\rm c}$ as a function of $n$ into   the McMillan equation and solve it inversely for  $\mu^*$.
It shows that $\mu^*$ decreases with $n$ and  the value  seems to be even negative   under $ \sim 10^{21} {\rm cm^{-3}} $. 
In their work, however,  the cause of this reduction is not discussed  beyond phenomenological analysis.

This result  is very curious,  because  $\mu^*$ represents the effect of repulsion as the screened Coulomb potential  and  should be positive.
Furthermore,  Eqs.\ref{mu_s_Morel} and \ref{mu_Morel} indicate that $\mu^*$  monotonically increases with the decrease of  $n$.\cite{mu-n-depend}
Therefore, it is hard to understand the behavior of  $\mu^*$, and  the result might seem  suspicious.
In this letter, we reexamine this problem and  clarify the puzzling behavior of  $\mu^*$  based on the plasmon mechanism  which has been  discussed on the superconductivity of  the uniform dilute electron gas.\cite{Takada1,Takada2,Rietschel}
It would give a reasonable  interpretation for $n$-dependence of   $\mu^*$ and recover  the usefulness  of  the McMillan formulation applying to the superconductivity of semiconductors.
%
%
\begin{figure}[b]
\begin{center}
\includegraphics[width=0.9 \linewidth]{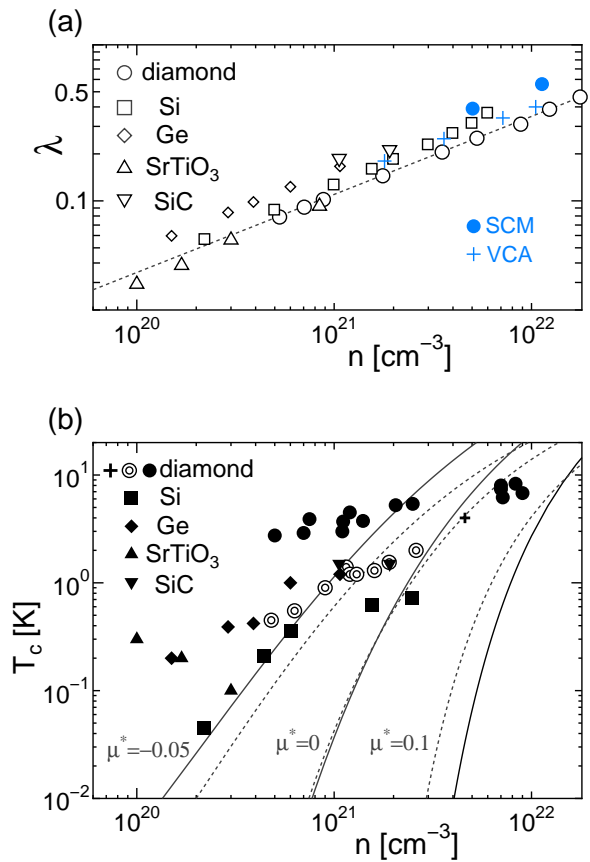}
\end{center}
\caption{(Color online)
(a) Theoretical estimation of  $\lambda$ by the RBA for diamond(empty circles), Si(empty squares), Ge(empty rhombus), SrTiO$_3$(empty triangle), and SiC(empty down-pointing triangle) with the results of the SCA(solid circles)\cite{Xiang} and the VCA(crosses)\cite{Ma} for diamond.
A doted line indicates a  fitting line to the result represented by empty circles. 
(b) Experimental $T_{\rm c}$ of diamond(cross\cite{Ekimov}, double circles,\cite{Klein} and solid circles\cite{Kawano}), Si(solid squares\cite{Grockowiak}), Ge(solid rhombus\cite{Herrmannsdorfer,Herrmannsdorfer2}), SrTiO$_3$(solid triangle\cite{Koonce}),  and silicon carbide(solid down-pointing triangle\cite{Kriener}).  
Solid lines and dotted lines indicate theoretical estimations of $T_{\rm c}$ of  diamond and  Si, respectively, where $\mu^*=-0.05, 0.0,$ and 0.1.
}
\label{fig1}
\end{figure}

At first,  we  clarify the discrepancy  between  results of experiments and  that of McMillan equation with a fixed $\mu^*$.
To do it, we calculate $n$-dependence of  $\lambda$  and $T_{\rm c}$ for several semiconductors, 
and  compare $T_{\rm c}$ systematically.
Calculations are performed using the 'Quantum ESPRESSO', which is a useful  computer code of the pseudopotential method.\cite{QE}  
We adopt the rigid-band approximation(RBA)\cite{Subedi}, which  is  the method  to introduce   fictitious carrier in the pure system by assuming the rigid-band. It changes only the Fermi energy of the system according to carrier density.  Although it is a simple approximation, it allows us to obtain systematic results for  any  semiconductors at any carrier density.\cite{mesh}    

In Fig.\ref{fig1}a, we show the result  of  $\lambda$ for diamond(empty circles), Si(empty squares), germanium(empty rhombus), strontium titanate(empty triangle), and silicon carbide(empty down-pointing triangle). Here we also show the results of the super-cell approximation(SCA)\cite{Xiang} and the virtual crystal approximation(VCA)\cite{Ma}, which are represented by solid circles and cross, respectively. Further, doted line indicates a fitting for the result of  diamond, which is given  by $ 0.11\times (n/10^{21})^{0.5}$. 
In the case of diamond,  $\lambda$ of the SCA\cite{Xiang} is  larger than  that of the RBA as shown in the figure.
Since the SCA  produces virtual periodicity of dopant in the system,   the value of $N(E_{\rm F})$ and $\lambda$ may be overestimated.\cite{Shirakawa,Yanase} 

On the other hand,  the result of  the VCA\cite{Ma} is almost consistent with that of the RBA. Indeed,  both methods should be equivalent to each other in the limit $n \to 0$.
In the RBA,  the effect of the impurity potential as a dopant  is completely neglected except  the effect of carrier doping.
Then,  the value of  $\lambda$ may be underestimated, especially in the diamond system.\cite{Yanase} 
However, if we attention to $n$-dependence of  $\lambda$, we find that these methods  give similar results.
Further,  our systematic calculation  indicates that $n$-dependence  is approximately described by a power function with respect to $n$, and the value of its power is near a half for all semiconductors within our calculation. 
 For example,  the powers of  Si and Ge are given by 0.56 and 0.52, respectively, which are close to that of diamond.
Because  semiconductors  have   similar  band structures  near $E_{\rm F}$ in the case of hole doping,   these results  may be plausible.\cite{power}

In Fig.\ref{fig1}(b), we compare experimental   $T_{\rm c}$  with  theoretical one.
Symbols indicate   $T_{\rm c}$ obtained by the experiments for  diamond(cross\cite{Ekimov}, double circles\cite{Klein} and solid circles\cite{Kawano}), Si(empty squares\cite{Grockowiak}), Ge(empty rhombus\cite{Herrmannsdorfer2}), strontium titanate(empty triangle\cite{Koonce}),  and silicon carbide(empty down-pointing triangle\cite{Kriener}). 
Solid lines represent the theoretical result calculated by our analysis  for  diamond and  doted lines are that of Si with $\mu^*=-0.05, 0,$  and 0.1.
Here, to calculate  $T_{\rm c}$ as a function of $n$, we use the  power function discussed in the above.  
 We also use $\omega_{\rm log}$ as 1600K for diamond and 430K for Si in  Eq.\ref{M_Tc}, which  are typical values and $n$-dependences of these are neglected for simplicity.
As shown in Fig.\ref{fig1}(b),  it indicates that  $n$-dependences of $T_{\rm c}$ obtained by theory  are far from that of  experiments, even if we use  negative $\mu^*$.\cite{Bustarret}
As long as a constant  $\mu^*$ is adopted to  Eq.\ref{M_Tc}, it is difficult to explain   $n$-dependences of $T_{\rm c}$. 
This result suggests that  we need to change the traditional interpretation  of $\mu^*$.

\begin{figure}[bt]
\begin{center}
\includegraphics[width=0.9 \linewidth]{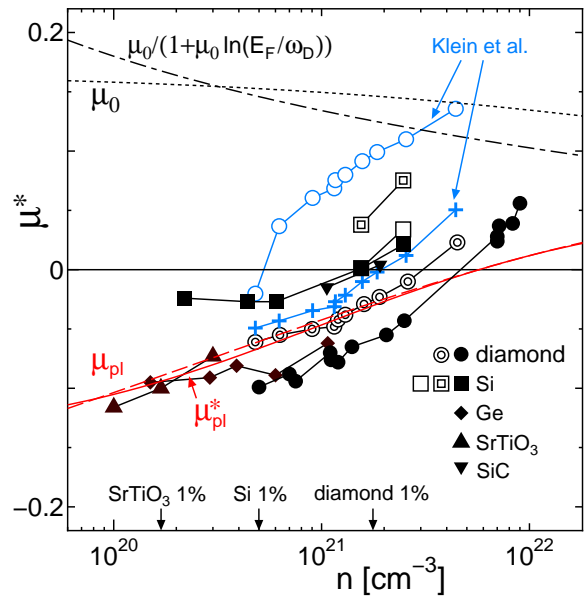}
\end{center}
\caption{(Color online) The Coulomb pseudopotential  $\mu^*$ using  experimental  $T_{\rm c}$ and $\lambda$ of the RBA, where symbols represent the result of diamond (double circles,\cite{Klein} and  solid circles\cite{Kawano}), Si(solid squares\cite{Grockowiak}), Ge(solid rhombus\cite{Herrmannsdorfer,Herrmannsdorfer2}), SrTiO$_3$(solid triangle\cite{Koonce}), and SiC(solid down-pointing triangle\cite{Kriener}). Further,  empty circles and crosses indicate the result of Klein. et al.\cite{Klein} for diamond, where the former  show the result  of  the SCA and the later are that of the VCA.
 Empty   and double squares indicate  the results\cite{Grockowiak}  of   the SCA and  the VCA for Si, respectively.
Solid, dashed,  dotted, and  dash dotted lines  represent  $\mu_{\rm pl}^*$,   $\mu_{\rm pl}$,  $\mu_0$,  and $\mu_0/(1+\mu_0\ln(E_{\rm F}/\omega_{\rm D}))$, respectively, where  $\omega_{\rm D}=2000$K.
Allows near border of the figure indicate the points corresponding to  the  doping concentration of 1\%
 for each semiconductor.}
\label{fig2}
\end{figure}

To focus on the behavior  of  $\mu^*$,  we  solve the McMillan equation inversely for $\mu^*$,
\begin{align}
\mu^*  \simeq \frac{1.04(1+\lambda)}{(1+0.62\lambda)\ln(1.2T_{\rm c}/\omega_{\rm  log})}+\frac{\lambda}{1+0.62\lambda}.
\label{mu_inv}
\end{align} 
Substituting  experimental $T_c$ and   $\lambda$  of the RBA to the above equation, we estimate  $\mu^*$ of several semiconductors.\cite{wlog}
In Fig.\ref{fig2}, we show  $\mu^*$ as a function of $n$ with the result of Klein et al.\cite{Klein}. 
They used $\lambda$  obtained  by the SCA  and the VCA to calculate $\mu^*$.
In the case of diamond,  $\lambda$ of the former  is larger than that of the latter. Therefore  the line of $\mu^{*}$ by the SCA(open circles)  is on the upper side than that of the VCA(crosses) and  the RBA(solid circles) in the figure.
Because, experimental $T_{\rm c}$ of the ref. [3]   is smaller  than that of  ref. [4],  it causes the difference between the results  represented by double circles and  that of  solid circles.
Figure \ref{fig2} clearly indicates that  $\mu^*$ is not a constant but a parameter depending on $n$.
It decreases with $n$, and  becomes negative in the case of $n \simk 3\times10^{21} {\rm cm^{-3}}$. 
Our result  also suggests that   the behavior of $\mu^*$ is not limited in diamond, but   common in  all semiconductors. 

In the above,  we show $n$-dependence  of $\mu^*$  obtained by the phenomenological method using the experimental  $T_{\rm c}$.
To explain this behavior theoretically, we obtain $\mu^*$  based on the plasmon mechanism  and compare it with the above result.
Then, we consider the gap equation of  the uniform electron gas system based on the work of Takada\cite{Takada1}, 
\begin{align}
\phi (x) =- \int_{-1}^{\infty} \frac{dx'}{2x'} \tanh(E_{\rm F} x'/2T_{\rm c})\phi (x') K(x,x'),
\label{phi}
\end{align}    
where $\phi (x)$ is a renormalized gap function defined as $\phi (0)=1$. Here, the kernel $K(x,x')$ is given by
\begin{align}
& K(x,x')=\frac{\sqrt{(1+x')} }{4g_{\rm  \nu} }  \nonumber \\
&  \times \int_{-1}^{1} dt \frac{ \omega_{\rm p} + (q_{\rm TF}^{2}/q^{2})(1+q^{2}/q_{\rm TF}^{2})^{1/2} (|x|+|x'|) }{(1+q^{2}/q_{\rm TF}^{2})^{1/2} \{ \omega_{\rm p}(1+q^{2}/q_{\rm TF}^{2})^{1/2}+ (|x|+|x'|) \} },
\label{Kxx}
\end{align}    
where  $g_{\rm  \nu}$ is valley degeneracy of the band, the frequency of plasmon $\omega_{\rm p}$ is given as $\sqrt{4\pi {\rm e}^2 n/m^{*}}$, $q^{2}$ is defined by $2m^{*}E_{\rm F}\{ 2+x+x'-2t\sqrt{(1+x)(1+x')} \} $, and   $q_{\rm TF}^{2}=4g_{\rm  \nu} {\rm e}^2 m^{*} p_{\rm F}/\pi$, with the effective mass $m^{*}$.
Here,  the dielectric constant is set to be unity for simplicity. 

Using $K(x,x')$ and $\phi (x)$,  $T_{\rm c}$  is given by 
\begin{align}
T_{\rm c}=1.134E_{\rm F} \exp(\frac{1}{ \mu_{\rm pl} }),
\label{Tc_pl}
\end{align} 
where 
\begin{align}
\frac{1}{\mu_{\rm pl}}=\frac{1}{\mu_0} + \int_{-1}^{\infty}\frac{dx}{2|x|}[\phi (x) K(0,x)/\mu_0-\theta(1-|x|)].
\end{align} 
Here,  $\mu_0=K(0,0)$ and $\phi (x)$ is  determined by 
\begin{align}
& \phi (x)=K(x,0)/\mu_0  \nonumber \\
& -\int_{-1}^{\infty}\frac{dx'}{2|x'|}\phi (x') [ K(x,x')-K(x,0)K(0,x')/\mu_0)].
\end{align} 
We noted that  $\mu_0$ is corresponding to $\mu$ of  Eq.\ref{mu_Morel} with  valley degeneracy  $g_{\rm  \nu}$.\cite{Takada1}. Thus, $\mu_{\rm pl}$ can be regarded as  an  effective Coulomb potential including  the effect of plasmon.

%
%

Of cause, the above analysis  is driven to the uniform electron system and  there is no energy scale of phonon.
So, we  introduce a fictitious phonon frequency $\omega_{\rm fi}$   to  Eq.\ref{Tc_pl} as
\begin{align}    
T_{\rm  c}=\frac{\omega_{\rm  fi}}{1.2} \exp(\frac{1.04}{ \mu_{\rm pl}^{*}}),
\label{Tc_pl2}
\end{align} 
where
%
\begin{align}
  \mu_{\rm pl}^{*}=\frac{1.04\mu_{\rm pl}}{1+\mu_{\rm pl} \log(1.3608E_{\rm F}/\omega_{\rm  fi})}.
\label{mu_pl}
\end{align} 
When we set $\omega_{\rm  fi}$ to $\omega_{\rm  log}$,  it  is completely corresponding to the McMillan equation  in the case of $\lambda \to 0$.
By replacing   $\mu^{*}$ with $\mu_{\rm pl}^{*}$,
we can expect that  Eq.\ref{M_Tc} gives  more  reasonable  $T_{\rm c}$ including the effect of plasmon   in the case of   $\lambda << 1$.
To confirm above consideration, we solve the above gap  equation numerically and calculate  $\mu_{\rm pl}$ as a function of $n$.
In Fig.\ref{fig2}, we show  $\mu_{\rm pl}$ and $\mu_{\rm pl}^{*}$,  where $\omega_{\rm  fi}$ is set as 1600K.
Here, the valley degeneracy and the effective mass of carrier are set as 3 and 0.5, respectively, to adapt our analysis to a typical  semiconductor.\cite{e-mass}
In the present systems, the value of $E_{\rm F}/\omega_{\rm fi}$ is not so large\cite{adiabatic}  and
then, the difference between  $\mu_{\rm pl}$ and $\mu_{\rm pl}^{*}$ is small.

 As shown in Fig.\ref{fig2}, the  $n$-dependence of $\mu_{\rm pl}^{*}$ is well consistent with that of  $\mu^{*}$ calculated by the experimental $T_{\rm c}$.  
It suggests that the McMillan equation replacing $\mu^{*}$ to $\mu_{\rm pl}^{*}$  is valid to analyze the superconductivity of not only for diamond but also for other semiconductors. 
In Fig.\ref{fig2}, the dash dotted line  represents   $\mu_0/(1+\mu_0\ln(E_{\rm F}/\omega_{\rm D}))$,  which is corresponding to the traditional $\mu^{*}$ without the effect of  plasmon.
It suggests that the effect of  plasmon reduces  $\mu^{*}$ to  about 0.2 at   $n \simeq 1.0 \times 10^{21}[{\rm cm^{-3}}]$, and its contribution  is fairly large as well as  phonon.
For  $n \simj 0.2 \times 10^{21}[{\rm cm^{-3}}]$,  $\lambda$ is smaller than  0.05 as shown in Fig.\ref{fig1}.
In this region,   the absolute value of $\mu^{*}$ becomes larger than  $\lambda$ and the  mechanism  of plasmon may be superior to that of phonon  to rise  $T_{\rm c}$.   
In fact,   $T_{\rm c}$ of  strontium titanate increases with the decrease of $n$, as shown in Fig.\ref{fig1}, although $\lambda$  decreases with $n$.
It indicates that  the effect of  plasmon dominates the superconductivity in this region.\cite{Takada2}
After all, the overall behavior of  $n$-dependence of  $\mu^{*}$ shows that  the mechanism of the superconductivity  is characterized by the interplay between   plasmon and  phonon.\cite{Hanke}

%

In summary, we  examine  the  Coulomb pseudopotential $\mu^*$  on the McMillan formulation in the case of  applying to the  heavily doped  semiconductors.
Systematic calculation of  the electron-phonon coupling constant $\lambda$  by the first-principles calculation  indicates that  $\mu^*$  should be not  a constant, but a  variable  depending on  carrier density $n$.
The  $n-$dependence of  $\mu^*$  indicates  that it decreases with $n$ and it becomes   negative under  $n \sim 2 \times 10^{-21}[{\rm cm^{-3}}]$. 
We find that the reduction of  $\mu^*$ is interpreted by the effect of  plasmon  which  may play an important role in the  superconductivity of  low carrier systems.
The effect   is fairly large and seems to be beyond the effect of  phonon for $n \simk 1 \times 10^{-21}[{\rm cm^{-3}}]$. 
Although our analysis is limited in the case of $\lambda << 1$, it  expands the validity of  the McMillan equation  into the superconductivity of semiconductors and explains the $n$-dependence of  $T_{\rm c}$ very well  in the case of $n \simk 5 \times 10^{-21}[{\rm cm^{-3}}]$. 

More sophisticated analysis including  the plasmon  effect has been  already  studied  on the superconductivity of lithium.\cite{Akashi}
However, the formulation is rather complicated and it may be hard task to apply to  the superconductivity of  semiconductors systematically.
Although the McMillan equation is more phenomenological method, it is widely accepted and  useful to estimate $T_{\rm c}$ of many materials.
We think that the reinterpretation of $\mu^*$ discussed in our analysis develops practical usefulness of the McMillan equation much further.

\section*{ACKNOWLEDGMENTS}

This work was supported by JPSJ KAKENHI Grant Number 15K05168.




\begin{references}
\bibitem{mesh}
In our calculation, we mainly use $20 \times 20 \times 20$  Monkhorst-Pack grid for the electronic Brillouin Zone integration and   $10 \times 10 \times 10$ mesh  for phonon calculation.
\bibitem{mu-n-depend}
In the electron gas, the $n$-dependence of $q_{\rm TF}^2/p_{\rm F}^2$ is given by $n^{-1/3}$.
\bibitem{wlog}
Here, $\omega_{\rm log}$  is  set as 250K, 510K, and 750K for Ge, SrTiO$_3$, and SiC, respectively.
\bibitem{power}
It is interesting to discuss the relationship between  $\lambda$ and  $N(E)$  based on the relation $\lambda \simeq VN(E_{\rm F})$.
 Because $V$ may be almost a constant with respect to  $n$, as demonstrated for picene\cite{Subedi} and diamond\cite{Sano},  the $n$-dependence of   $\lambda$ may be determined by  that of $N(E_{\rm F}) $.
If we assume   $N(E)  \propto   E^{\beta} $ near $E_{\rm F}$, we get $N(E_{\rm F})  \propto   n^{\frac{\beta}{1+\beta}} $ by using the relation  $n=\int_{E_{\rm F}}^{0}  N(E)  dE $ in the hole doping case. 
The result of the RBA(not shown) indicates that  $\beta$ is  about unity for diamond. It leads  $\lambda \propto  n^{0.5}$ as shown in  Fig.\ref{fig1}.
\end{references}
\end{document}